\documentstyle[12pt]{article}
\begin{document}
\setlength{\baselineskip}{24pt}

\begin{center}
{\large VARIATIONAL MONTE CARLO AND CONFIGURATION INTERACTION STUDIES
OF $C_{60}$ AND ITS FRAGMENTS}\footnote{Contribution no. 1106   from the Solid State
and Structural Chemistry Unit} \\
\vspace{1cm}
Bhargavi Srinivasan$^{2}$, S. Ramasesha$^{2,4}$ and H. R.
Krishnamurthy$^{3,4}$  \\
\vspace{0.5cm}
$^{2}$Solid State and Structural Chemistry Unit  \\
Indian Institute of Science, Bangalore 560 012, India \\
   
\vspace{0.5cm}
$^{3}$Department of Physics  \\
Indian Institute of Science, Bangalore 560 012, India \\   
\vspace{0.5cm} 
$^{4}$Jawaharlal Nehru Centre for Advanced Scientific Research \\
Jakkur Campus, Bangalore 560 064, India \\
\end{center}
\pagebreak
\clearpage
\begin{center}
{\bf \underline{Abstract}}\\
\end{center}
The  $C_{60}$ molecule and its fragments  are studied
using  Configuration Interaction (CI) and Variational Monte Carlo
(VMC) techniques, within the Hubbard model. Using benzene as a test
case, we compare the results of the approximate calculations with
exact calculations. The  fragments of $C_{60}$ studied are
 pyracylene, fluoranthene and corannulene. The
energies, bond orders, spin--spin and charge--correlation functions
of these systems are obtained for various values of the Hubbard parameter, $U$. 
The analysis of bond orders and correlation functions  of these individual molecules allow us to
visualise  pyracylene as a naphthalene unit with two
ethylenic moieties and fluoranthene as  weakly bridged benzene and
naphthalene units.  Corannulene is the largest fragment of
$C_{60}$ that we have studied.  The hexagon--hexagon(h--h) bond orders are slightly larger than those  of the
hexagon--pentagon bonds(h--p), a feature also found in other fragments. 
We also find bonds between two coordinated carbon sites
to be stronger than bonds involving three coordinated carbon sites.
In $C_{60}$, the h--h bonds are stronger than in corannulene and the
h--p bonds weaker than in corannulene for all correlation strengths.
Introducing bond alternation in  the buckyball enhances this difference. 

\pagebreak
\section{Introduction}
The discovery of $C_{60}$, a new allotrope of carbon, by Kroto et al\cite{kroto}
and its bulk synthesis by Kratschmer et al\cite{Kw} have been amongst
the most exciting recent developments in chemistry and physics. $C_{60}$ and its compounds
show exotic physical and chemical properties. Alkali doped
compounds  of $C_{60}$ are superconducting, with $Rb_{1.5}Cs_{1.5}C_{60}$
 having a 
superconducting transition temperature
of 32 K\cite{fleming}. The compound $C_{60}-TDAE$
($TDAE$=tetrakis dimethyl amino ethylene) is an organic ferromagnet
with a fairly large Curie temperature of 16.5K\cite{alle,steph}. 

To develop models for diverse phenomena such as these, it is essential
to have a reliable understanding of the electronic structure of 
the basic building blocks of these systems. The carbon atoms in
fullerenes are in a nearly $sp^{2}$ hybrid configuration with a
half--filled $p_{\pi}$--like orbital directed radially and involved in extended conjugation.
Electronic structure studies on conjugated  organic systems have
brought out clearly the importance of electron correlations to
account for many of the observed spectroscopic and related
properties\cite{srjcp}.
Explicitly including electron correlations is a formidable challenge even
at the level of a single buckyball. It is reasonable to hope that a proper study
of a single buckyball and its fragments will shed  light on
 the esoteric electronic properties of these
systems in the bulk. 

It would be interesting to try and understand
the properties of the full buckyball in terms of its fragments.
 Firstly, for small enough fragments, it is
possible to perform exact calculations that allow us to
systematically check the approximate methods. The
electronic structures of the fragments are expected to be useful in interpreting the
electronic structure and geometry of $C_{60}$.
 In this article, we study the ground state properties such as bond
orders, charge--charge and spin--spin correlation functions 
of some fragments of $C_{60}$ using configuration interaction (CI)
  techniques and the Variational Monte Carlo (VMC)
method.  We report the results of our  VMC calculations on
the full buckyball within the Hubbard model, with and without
bond-alternation.

The simplest description of the electronic structure of $C_{60}$ is
found within the framework of the 
Huckel picture, wherein  a non-zero transfer integral is
introduced between the $p_{\pi}$--like orbitals on nearest--neighbour
atoms(sites).  The resulting molecular orbitals (MOs) have a 
high degree of degeneracy due to the icosahedral symmetry of the
molecule. Neutral $C_{60}$ has a five-fold degenerate set of highest
occcupied molecular orbitals (HOMOs) and a three-fold degenerate set
of lowest unoccupied molecular orbitals (LUMOs). The high degeneracy of
the MOs has been the focal point of the discussion of the electronic structure
of these systems. For instance,  in a first approximation, the
degeneracy  of the LUMOs is considered essential to rationalize
the existence of ferromagnetic exchange between two monoanions of
$C_{60}$\cite{bssrfst}.

The minimal interacting description for $C_{60}$ is given by the single band
Hubbard model which incorporates only an on-site electron--electron repulsion 
$U$ besides
the one-electron part given by the Huckel picture. Drastic
truncation of  the electron repulsion
to an on--site interaction is justifiable in a metal where the
mobile conduction electrons lead to short Debye--Huckel screening lengths.
In molecules it is strictly necessary to include extended range interactions,
at least at the level of the PPP model, because of inefficient screening.
However,  obtaining reliable results even for such 
models is difficult and we study the simpler Hubbard model as 
a first step in understanding crucial {\it qualitative} effects of
correlations in molecular systems like $C_{60}$ and its fragments.
Exact diagonalization techniques are useful for
small systems (with less than about 14 sites at present) but are impractical for large systems
because of the exploding  dimension of
the basis with increasing system size\cite{sranthra}. Therefore, approximate
approaches to the problem are inescapable for larger systems.
However, exact diagonalization provides a strong check on any novel
approximation scheme.

Amongst approximate techniques, perturbation theory gives fairly reliable 
results at limiting values of the model parameters but breaks down for
intermediate but relevant values of the
parameters\cite{sudip}.  Another equally widely used approximate  approach has been
the variational approach with the Gutzwiller wavefunction (GWF) as the trial
function of choice for Hubbard models\cite{gutz}. In this approach, a
variational parameter $g$ ( $0 \le g \le 1$) is introduced  and the
weight of a configuration in the site representation of the
Hartree--Fock  ground state is modified by a factor $g^{D}$, where
$D$ is the number of double occupancies in that configuration.
Even though a single variational parameter $(g)$ is to be determined, the
method itself is not easy to implement. Evaluation of the expectation
value of the Hamiltonian for the GWF is made difficult by the very
large number of configurations appearing in the wavefunction in the
site representation. Often, uncontrolled approximations are resorted
to for evaluating the expectation value\cite{gutz,joyes}. Such
approximations can be avoided if we carry out  a Monte Carlo
integration  for the expectation value,
using the VMC method. The VMC technique uses
an  importance sampling of the various Slater Determinants (in the
atomic orbital (AO) basis) that make up the
Gutzwiller trial function to estimate the expectation value of the
Hamiltonian.

 We also employ  a novel
configuration interaction (CI) technique that
considers all many-body states derived from multiple particle--hole
excitations from the Hartree-Fock ground state, whose one 
particle energies lie below a chosen cutoff. The technique gives
reliable ground and excited  states of small fragments of $C_{60}$ for
comparison with Monte Carlo studies. We carry out  VMC  studies
on the full buckyball using the optimized  Monte Carlo  parameters
arrived at from such comparisons.

This article is organized as follows. In the next section we give a
description of the computational schemes we have employed, namely the
approximate CI scheme and the VMC method employing Gutzwiller trial
function. We provide detailed comparisons of these techniques with
exact results for benzene. Section 3 deals with our results and discussions
on fragments of $C_{60}$. In section 4 we discuss our VMC results
on $C_{60}$ and compare these with results on fragments. 
The last section summarizes the main results of the paper.

\section{Computational Schemes}

The computational schemes that are described in this section are
applied to the Hubbard model\cite{hub} on the buckyball and its fragments. The
Hubbard Hamiltonian may be written as
\begin{equation}
 H_{Hub} = H_{0} + H_{1} 
 \end{equation}
 \begin{equation}
  H_{0} =  \sum_{<ij>}\sum_{\sigma}t_{ij}
(a_{i\sigma}^{\dagger}a_{j\sigma}+a_{j\sigma}^{\dagger}a_{i\sigma})
\end{equation}
\begin{equation}
 H_{1} = U\sum_{i}\hat{n}_{i \uparrow}\hat{n}_{i \downarrow}\
\end{equation}
\noindent          
where $H_{0}$ is the non-interacting part of the Hamiltonian and $H_{1}$
represents the on-site Coulomb repulsion. The operator
 $a^{\dagger}_{i\sigma}$ ($a_{i\sigma}$) creates (annihilates) an
electron with spin ${\sigma}$ in the orbital at the $i^{th}$ 
site, $\hat{n}$$_{i\sigma}$ are the corresponding  number operators
 and the summation $<ij>$ is over bonded atom pairs. The transfer
integral in all the computations is fixed at 1.0 and energies are
reported in units of $t$ unless specified otherwise.

\subsection{VB Method in the MO Basis}
In carrying out approximate electronic structure calculations of large
systems, within a restricted CI scheme, it is important to find a 
criterion for truncating the basis. In general, it is difficult to find 
a truncation scheme to restrict the size of the Hilbert space, that
yields accurate spectral gaps. The crux of the problem is that
increasing the dimensionality of the basis set in the restricted CI
scheme undoubtedly improves the ground state energy in accordance
with the variational theorem, but does not guarantee a concommitant
improvement in the excited state energies.  The scheme described here
is  intuitive and does not have the 
drawback of the uncontrolled nature of the multi-reference CI (MRCI) 
schemes in which the MOs chosen in the reference determinants are
arbitrary\cite{szabo}.

In our restricted CI scheme,
we  employ the MOs to construct the VB functions\cite{bssr}. 
The  Hubbard  Hamiltonian in the MO basis can be written as
\begin{equation}
H = \sum_{p \sigma} \epsilon_{p}b_{p \sigma}^{\dagger}b_{p \sigma} + 1/2
\sum_{pqrs} W_{pqrs}(\hat{E}_{pq} \hat{E}_{rs} - \delta_{qr}\hat{E}_{ps}) 
\end{equation}
\begin{equation}
\hat{E}_{pq} = \sum_{\sigma} b_{p \sigma}^{\dagger} b_{q \sigma}
\end{equation} 
where  $b_{p \sigma}^{\dagger}$ ($b_{p \sigma}$) corresponds to creation
(annihilation) of an electron in the $p^{th}$ molecular orbital with
spin $\sigma$, $\epsilon_{p}$ is the energy of the $p^{th}$ MO.
$W_{pqrs}$ is defined by
\begin{equation}
W_{pqrs} = U \sum_{i } c_{ip} c_{iq} c_{ir} c_{is}
\end{equation}
\noindent
where $c_{ip}$ is the coefficient of the $i^{th}$ AO in an AO
expansion of the 
$p^{th}$ MO and all the summations run over the entire respective bases.
 The Hubbard Hamiltonian written in the MO basis 
has the property that the one-electron part is diagonal, while the interaction
part is off-diagonal. We use the Rumer-Pauling\cite{soos} rules to construct the
VB diagrams from the MOs. However, if an MO state is degenerate, the
associated MO functions may not be orthogonal. To avoid the 'nonorthogonality
catastrophe', we construct the  Lowdin orthogonalized MOs\cite{low} whose
coefficients are given by
\begin{equation}
{\bf c}^{L} = {\bf S}^{-1/2} {\bf c}
\end{equation}
\noindent
where $\bf{S}$ is the overlap matrix in the MO basis and
$\bf{c}$ and ${\bf{c}}^{L}$ are the
coefficient matrices. We then form the
matrix representation of the Hamiltonian in the MO--VB basis using rules
similar to those developed for AO--VB calculations\cite{srjms,srzgs}.
The restricted MO--VB basis  includes all configurations
whose energy with respect to the energy of the ground state
configuration is below a given threshold, $E_{c}$.  We have observed
that when $E_{c}$ equals the spread in the one--particle spectrum   
(bandwidth of the extended systems), the excitation gaps smoothly follow
the exact gaps in small systems. The limitation of this approach
arises from the fact that as the system size increases, the Hamiltonian
matrix becomes less sparse and this places a restriction on the
largest system size accessible by this technique. If we are only
interested  in the ground state properties of the system, keeping the
largest possible cutoff would be appropriate in accordance with the
variational principle. We usually keep $E_{c}
\approx$  bandwidth for ground state properties of a single molecule.

\subsection{Variational Monte Carlo with the Gutzwiller Trial Function}

 The VMC method employs a trial wavefunction, usually
of the Jastrow form\cite{bijl,jas} and provides a Monte Carlo estimate of the
expectation value of the Hamiltonian and other operators in this
state for different values of the variational parameters.
Here, we focus on the Gutzwiller trial wavefunction which is the
simplest  Jastrow function and is appropriate for Hubbard
models. In this
section, we briefly describe our implementation of the VMC method,
which is similar to that of Yokoyama and Shiba\cite{yok}.

The Gutzwiller wavefunction is given by 
\begin{equation}
 |\Psi_{G}> = g^{\hat{D}} |\Phi> 
 \end{equation}
 \begin{equation}
 \hat{D} = \sum\limits_{i} \hat{D}_{i}
 \end{equation}
 \begin{equation}
 \hat{D}_{i} = n_{i\uparrow} n_{i\downarrow}
 \end{equation}
 \noindent
 where $g$ is the sole variational parameter. Here $|\Phi>$ is the ground state  
, given by,
\begin{equation}
|\Phi> = |\Phi_{\uparrow}>|\Phi_{\downarrow}>
\end{equation}
\begin{equation}
|\Phi_{\sigma}> = \prod\limits_{p}b^{\dagger}_{p\sigma}|0>
\end{equation}
\noindent
where the product is over the MO's occupied
with electrons of spin $\sigma$. The state $|\Phi>$ can be written in terms of
the states $|R>$, where $|R>=|R^{\uparrow}>
|R^{\downarrow}>$ represents a given configuration of up
and  down spins in real space (equivalently in AO or site
representation), as
 \begin{equation}
 |\Phi>=\sum\limits_{R^{\uparrow}}\sum\limits_{R^{\downarrow}}
 det ({\bf{C}}^{\uparrow}(R_{n}^{\uparrow}))
 det ({\bf{C}}^{\downarrow}(R_{n}^{\downarrow})) 
 |R^{\uparrow}> |R^{\downarrow}>.
 \end{equation}
 \noindent
Here  $det ({\bf{C}}^{\uparrow}(R_{n}^{\uparrow}))$
is an $N_{\uparrow}
\times N_{\uparrow} $ determinant  whose
elements in the $j^{th}$
row correspond to the coefficients of the $j^{th}$ occupied MO (for
this spin) at the sites occupied in the configuration $|R_{n}^{\uparrow}>$.
  The operator  $\hat D$ picks out
the number of doubly occupied sites $d(R)$ in every configuration $|R>$ appearing in
$|\Phi>$ (eqn. 13).
 
 In the VMC method, we seek to evaluate the energy expectation value
of the Hamiltonian in the state $|\Psi_{G}>$,
\begin{equation}
 E(g)  = { {<\Psi_{G} | H | \Psi_{G}>} \over  {<\Psi_{G} | \Psi_{G}>}}
 \end{equation}
 \noindent
To do so, we rewrite $E(g)$ as
\begin{equation}
 E(g) = \sum\limits_{R} P_{g}(R) {{<\Psi_{G}|H|R>} \over {<\Psi_{G}|R>}}
 \end{equation}
 \noindent
where $P_{g}(R)$ is the probabilty of occurance of the configuration
$|R>$  given by
\begin{equation}
P_{g}(R) ={ { |<R|\Psi_{G}> |^{2} } \over {\sum\limits_{R^{\prime}} 
|<R^{\prime}|\Psi_{G}> |^{2}}}.
\end{equation}
\noindent
From eqn. (13) for $|\Phi>$, it follows that 
 \begin{equation}
<R_{n}|\Psi_{G}>= g^{d(R_{n})}
 det ({\bf{C}}^{\uparrow}(R_{n}^{\uparrow}))
 det ({\bf{C}}^{\downarrow}(R_{n}^{\downarrow})) .
 \end{equation}
 \noindent

A direct importance sampling of the configurations  $\{|R>\}$ with weights given
by eqn. (16) is not possible since  there is an explosion in the number
of configurations for large systems and the denominator in eqn. (16)
cannot be evaluated explicitly. Therefore, we resort to an indirect
importance
sampling technique which is achieved by a Markov process. The states
of the Markov chain are the configurations and the one--step
transition probability between configurations is such as to yield
$P_{g}(R)$ (eqn. (16)) in the ergodic limit. In the computational
procedure, we generate the configuration  $|R_{n+1}>$ from $|R_{n}>$  by
hopping an electron of randomly chosen spin, from a randomly chosen occupied
site, to a randomly chosen vacant site. The new configuration
$|R_{n+1}>$ is accepted with probability 
\begin{equation}
 p_{n,n+1} = 
 { { | <R_{n+1}|\Psi_{G}>|^{2} } \over { |<R_{n}|\Psi_{G}> |^{2}}}
\end{equation}
\noindent

If the new configuration is not accepted, the configuration $|R_{n}>$ is
itself treated as the new configuration. The ratio $p_{n,n+1}$ is given
by
\begin{equation}
p_{n,n+1}   =   g^{2 \delta d}
{{| det ({\bf{C}}^{\uparrow}(R_{n+1}^{\uparrow})) |^{2} } \over {|det ({\bf{C}}^{\uparrow}(R_{n}^{\uparrow})) |^{2}}}
{{| det ({\bf{C}}^{\downarrow}(R_{n+1}^{\downarrow})) |^{2} } \over {|det ({\bf{C}}^{\downarrow}(R_{n}^{\downarrow})) |^{2}}}
\end{equation}
\noindent
where $\delta d ~( = d(R_{n+1})-d(R_{n}))$ denotes the change in the
number of doubly occupied sites brought about by the electron hop.
We use the updating algorithm of Ceperley et al\cite{cep} to compute the ratio
of determinants appearing in eqn. (19). This involves updating the
inverses of the matrices ${\bf{C}}^{\sigma}(R^{\sigma})$ which can be
carried out in 
$N^{2}$ arithmetic operations\cite{numrec}.

The energy expectation value is computed as a sum of the transfer
part and the interaction part. It is trivial to compute
the expectation value of the interaction part of the Hamiltonian,
because it only involves counting the number of doubly occupied sites
in a given state. The effect of the hopping operator $\Delta_{i}^{\sigma}$ 
on a state $|R_{n}^{\sigma}>$, that hops an electron from site $i$  is
\begin{equation}
\Delta_{i}^{\sigma} |R_{n}^{\sigma}> = \sum\limits_{j}
 |R_{n}^{\sigma, i \rightarrow j}>
\end{equation}
\noindent
where $|R_{n}^{\sigma, i \rightarrow j}>$ denotes a configuration
generated from $|R_{n}^{\sigma}>$ by hopping an electron of spin
$\sigma$ from site $i$ to site $j$.
 The non--interacting part of the Hamiltonian
is the sum of $\Delta_{i}^{\sigma}$ over all sites $i$ and spin $\sigma$. Thus,
 we can write the contribution of the state $|R_{n}>$ to the
total  energy expectation value for the Gutzwiller wavefunction
$\Psi_{G}$ as
\begin{equation}
 {{<\Psi_{G}| H |R_{n}>} \over {<\Psi_{G}|R_{n}>}}  =  -t \sum\limits_{i}
 \sum\limits_{\sigma}\sum\limits_{j}
{ { <\Psi_{G}|R_{n}^{\sigma, i \rightarrow j}>  } \over
{<\Psi_{G}|R_{n}^{\sigma}> }} + U N_{d}
 \end{equation}
 \noindent
 The calculation of other expectation
values is similar to the calculation of energy. In particular,
the expectation value of any operator $A$ can be estimated by estimating
 $<\Psi_{G} | A| R> / <\Psi_{G} | R> $. As
before, the averaging  is carried out after allowing sufficient
number of moves per electron (MPE) to reach the ergodic limit.

The standard deviation in the estimated quantities can be taken as 
an error estimate in any Monte Carlo procedure. In the estimates
of energies and related quantities, this requires computing the 
expectation value of $H^2$ where $H$ is the Hamiltonian. The number 
of resultant Slater Determinants when the Hamiltonian acts on a single
Slater Determinant is of $O(p)$, where $p$ is the number of transfer
terms in the Hamiltonian. In $C_{60}$, $p=360$ as there are 90 nearest
neighbour bonds. Therefore, in the estimation of the 
expectation value of $H^2$,
the number of Slater determinants resulting from the operation of $H^2$
on a single Slater determinant is of $O(p^2)$. Thus, computationally, 
it is very expensive to estimate the expectation value of $H^2$ and hence
the standard deviation.
However, small systems are
amenable to {\it exact} Gutzwiller calculations and we have  compared
VMC estimates with these results to arrive at a satisfactory number
of Monte Carlo steps. The actual scaling of Monte Carlo errors
with size of the physical system is not known. In large systems, 
for some model parameters, we have checked that the
Monte Carlo averages obtained with $10^{4}$ MPE, after allowing about 
2000 MPE for reaching ergodicity, do not change within the significant 
figures quoted in the tables even upon doubling the sample size.

\subsection{Comparison of Approximate and Exact Results for Benzene}
We first compare our results from the VMC method with exact
calculations in the case of benzene. The calculations on benzene
include computation of  expectation values obtained from the
variation of $g$ in the full Gutzwiller
wavefunction and also a Monte Carlo estimate of the same. We refer to
the former as the true Gutzwiller value and to the latter as the VMC
value. In  table(1), we
compare the energy from exact--diagonalization with   the true
Gutzwiller energy and the VMC  
as a function of the Hubbard parameter, $U$. We observe that
even upon increasing $U$, both the true expectation value  in the
Gutzwiller  wavefunction   and  the MC estimate continue to compare
 well with the exact energy. 
 
 The ground state energy alone is not sufficient for a  physical picture of the
system and  it is essential to have reliable correlation functions.
Therefore, we have studied the spin--spin and charge--charge
correlation functions using all the techniques mentioned above. 
In table(2), we present the spin--spin and charge--charge correlation functions obtained
from the exact calculation, the true Gutzwiller value and
the VMC estimate for weak and intermediate on--site correlation
strengths $U$. 
We observe that  the 
GWF gives  a good estimate of the correlation functions and that
the errors in the VMC are not much larger. However, it is worth
noting that the Gutzwiller method slightly underestimates the role of
correlation as is seen from the higher than exact values of
$<n_{i}n_{j}>$ and the lower than exact values of for
$|<s_{i}s_{j}>|$ for both $U=1.0$ and $U=4.0$. The reduction in charge
fluctauations and concommitant increase in spin fluctuations with
increase in $U$ is suggestive of the development of a spin at each
site in the large $U$ limit. 
In table (3), we present the exact, true Gutzwiller and VMC bond--orders for two different values of
$U$. We see that the VMC and the Gutzwiller bond--orders compare well
with the exact  values for wide range of correlation strengths. The
larger bond order in the Gutzwiller method compared to the exact
value is consistent with the observation that this approximation
slightly underestimates the role of $U$. 

The results presented in tables (2) and (3) are obtained after
averaging over equivalent site pairs and up and down spins. The VMC
estimates before averaging are not strictly equal for equivalent
pairs while the true Gutzwiller numbers reflect the symmetry of the
system.  

The comparison between MO--VB and exact values in the ground state
are  favourable for threshold $\approx  $ bandwidth,
although the MO--VB technique is designed to reproduce excitation
gaps for various correlation strengths and is not the method of
choice for the ground state properties. 

\section{Results on Fragments of $C_{60}$}

In this section, we present our results on the fragments of $C_{60}$
using the MO--VB technique and compare these with results obtained
from the VMC method.
We have studied some catacondensed fragments of $C_{60}$ which
contain characteristic chemical units of $C_{60}$. These fragments are
pyracylene,  fluoranthene and
corranulene, shown in fig(1). These fragments contain one or more
five--memebered rings  essential for displaying the minimal
topological features of $C_{60}$. The trends observed in these systems
are expected to be indicative of the behaviour of  the full molecule.

\subsection{Pyracylene}

The pyracylene molecule is a fragment of $C_{60}$ that is 
amenable to exact calculations. The study of this molecule allows a
further test on the approximations besides being of importance in its
own right as the molecule is a characteristic fragment of $C_{60}$. 
In table(4) we present the energy of pyracylene for various values of 
 $U$. The energy obtained from the VMC calculation compares  favourably with
that from the exact calculation for $U=1.0$.

  In table(5a) we present the bond order data for the unique bonds in
pyracylene. From the bond order data, we see that the 1-2 
bond or its equivalents is the strongest bond in the system. This trend is found for
all correlation strengths studied and is also seen in the exact PPP\cite{acyam}
ground state. The external carbon atoms of the pentagons form almost
independent  ethylenic units with a slight  extended conjugation as seen
from the rather small 2-3  bond order. This is summarized in fig. 2
for $U/t=1.0$. The bond
orders of the internal bonds of the six membered rings are smaller
than that of benzene for the same value of $U$ (table(3)) although we
can still view the electrons as being delocalized over the benzene
rings. This picture is reinforced by  the ring currents obtained in
the PPP  model which show large diamagnetic
circulations around the naphthalenic perimeter but paramagnetic
circulations around the pentagons. The similarity in PPP and Hubbard
bond orders implies similar ring currents in the Hubbard model also.
We  also expect paramagnetic circulations along the perimeter
as found in the PPP model\cite{acyam}. 

The kinetic energy contribution to the total ground state
energy  decreases with increasing $U$ and comparison with the PPP
results allows us to estimate that the Hubbard model with an
$U_{eff}$ of  nearly $3.0t$ should be similar to  the PPP
model.
In table(5b) we present the bond order data of table(5a) normalized
with respect to the 1-2 bond. Clearly, the 
Hubbard model shows smaller bond order variations compared to the PPP
model.  If the off--site interaction part in the PPP model is
treated in the mean field, it would renormalize the transfer part of
the Hamiltonian.  It is for  this reason that the interaction term in
the  PPP model
influences the geometry, while the Hubbard interactions, which are
purely on--site have a smaller influence on the geometry of the system. 
  
In tables(6a) and (6b) we present the spin and charge correlations for
both on--site and nearest neighbour atom pairs and summarize them
in fig. 2 for $U/t=1.0$. The 1-2 bond has a very
large antiferromagnetic spin correlation  (-0.402) although the
sites 1 and 2 have rather large ionic contributions. In the half-filled case,
the diagonal (on-site) spin and charge correlation functions in the 
large $U$ limit are both 1.0,  while  in the $U \rightarrow 0 $ limit, 
they are 0.5 and 1.5 respectively.\footnote{These results are obtained
by computing $4\langle s_{i}^{z}s_{i}^{z}\rangle$ and 
 $\langle n_{i}^{2}\rangle$, given that for the half-filled case,
in the $U/t=0$ limit, the probabilities of the Fock space states
$( |0\rangle, ~~|\uparrow \rangle, |\downarrow \rangle and 
|\uparrow \downarrow \rangle )$, at a given site $i$,
 are all equal to 0.25.  In the large $U$ limit, these probabilties
are  respectively 0, 0.5, 0.5 and 0. }
 The large value of $<\hat{n}_{1}^{2}>$
and $<s_{1}^{z}s_{2}^{z}>$ for the sites 1 and 2 can be reconciled 
only if we assume that
the spins are aligned antiparallel with a high probability whenever
these sites are singly occupied.  Thus the ethylenic units show no
frustration for delocalization, a trend also evident from  the bond
order.  The spin correlations follow the same trend as the bond
orders, with spins at sites 2 and 3 being almost uncorrelated. This
implies that the 2--3 bond is not frustrated. The
most ionic sites in the molecule are the interior sites, 4 and 11.
The exterior benzene sites are  least ionic . The
hexagon--hexagon bond is also found to be marginally stronger than
the hexagon--pentagon bond, a feature in common with $C_{60}$.

\subsection{Fluoranthene}
The dependance of the energy of  fluoranthene on $U$ is given in table(4).
Here, exact results are not available for
comparison but the energy scales as expected with $U$. 
From the bond order data of fluoranthene (fig. 3a), we see that the pentagon bonds are
weak, while the bonds on the perimeter are  strong.  The weakest
bond  is the bridging 4--5 bond. The bonds forming the benzenic unit
and the naphthalenic unit are fairly strong. From the bond order
data, we can visualize fluoranthene as  weakly bridged benzene and
naphthalene units, a feature consistent with an earlier PPP model
calculation. Furthermore,  the bond order in the isolated
hexagon is nearly the same as in benzene. The bonds in the
naphthalenic unit show more alternation. We notice that even in
fluoranthene, the hexagon--hexagon bond is slightly stronger than the
hexagon--pentagon bond.  The dependence of bond order on correlation
strength is shown in table (7).  All the bond orders decrease with
increasinng $U$. The decrease is sharper at larger $U$ values. The
bond orders tend to become uniform within the benzenic and
naphthalenic units. 

The spin and charge correlations of the nearest neighbour bonds for
$U=1.0$ are shown in fig(3b).
The bridging 4--5 bond has a very small antiferromagnetic spin
correlation, despite being a nearest neighbour bond. Thus we expect
most of the frustration of the pentagon to be manifest in this
bond.  The charge--charge correlation is largest for this atom pair,
and so is the on--site charge correlation. The small spin--spin
correlation   and the large charge correlations and site
ionicities of this bond can be reconciled if we assume that both the
sites are simultaneously empty or doubly occupied, and if singly
occupied the spins are uncorrelated. It also implies that the most
reactive sites in fluoranthene are likely to be these sites. In
tables  (8a) and (b) we present  the dependence of charge and spin
correlation son $U$. While all the sites tend to be more covalent,
the relative ionicities as well as charge and spin correlations
evolve smoothly with $U$. 
 We also observe that as $U$ increases, the bond orders decrease,
while the spin--spin correlations increase, a trend expected from the
Hubbard model.

\subsection{Corannulene}
The energy of the ground state of corannulene for different values of
$U$ is presented in table(4).  The ground state energy per carbon
atom, $\epsilon_{c}$ is the lowest for this fragment, with
$\epsilon_{c}$ for the fragments obeying the trend corannulene < fluoranthene < pyracylene.  The
stabilization decreases with increasing $U$, although the same trend
persists. In fig. (4), we present the the bond order data  of
corannulene for $U=1.0$. Unlike in the  other two fragments of
$C_{60}$,  the pentagonal unit in corannulene is
completely enclosed. Therefore the bonds of the pentagon in this
fragment most closely approximate the h--p bonds in
$C_{60}$.  All h--p bonds in corannulene  have almost equal bond orders.
 We observe that the h--p bonds are slightly weaker
than the h--h bonds. The outer bonds of corannulene also
fall into two categories. The 1--2 like bonds are 
stronger than the 2--3 like bonds. The higher
bond orders of the former group suggests that bonds involving even a
single three coordinated carbon atom tend to be weaker due to
frustrations  encountered during delocalization.
The outer bonds with  large bond orders 
evolve  into the h--h bonds in $C_{60}$. 
The bond orders show similar trends for different values of $U$
(table (9)), although the magnitudes decrease with increasing correlation
strength. 

The spin--spin and charge--charge correlations for $U=1.0$  are shown
in fig. (5). The interior  sites are more ionic than the exterior
sites.  The charge--charge correlations for atom pairs in the
interior sites is larger than for the atom pairs in the exterior
sites.  The charge--charge  correlation for the nearest neighbour
pair involving one interior and one exterior atom  is in between  the
values for the purely interior and the purely exterior atom pairs.  The smallest charge--charge
correlation is found for  bonds with the largest bond order. The
diagonal spin correlations are nearly uniform at all sites. The
spin--spin correlations are nearly the same for any atom pair
involving atleast one three connected carbon atom, while, for the atom
pair with both two--connected carbons, it is  much larger.  This
reflects the bond frustration in bonds involving three connected
carbon atoms.  These trends continue to hold  for larger values of
$U$, as seen from tables (10a) and (10b).

\section{VMC Ground State Properties of $C_{60}$}

We have obtained the ground--state energy of neutral $C_{60}$, with and
without bond--alternation for several values of $U$.  We have also
computed the nearest neighbour bond orders and the charge and spin
correlations of $C_{60}$ in both cases. 
The energies of $C_{60}$ are presented in table (11). These
energy values are in good agreement with those of Krivnov
et al for the values of $U$ studied\cite{kriv}. We note that
the energy per carbon atom is lower than that of any of the fragments
studied. The electronic energy is further reduced on introducing
bond alternation.  However, we need a reliable estimate of the strain
energy\cite{varma,schul} to be able to arrive at an equilibrium distortion of the bonds
in $C_{60}$. Nevertheless, it is now established that the h--h bonds in
$C_{60}$ are shorter than the h--p bonds\cite{wifd,hed}. It is interesting to
compare the energy per bond of the fragments with that of $C_{60}$.
The $U=1.0$ values for benzene, pyracylene, fluoranthene, corranulene
and $C_{60}$ are -1.10, -0.95, -0.99, -0.96 and -0.88 respectively.  This trend is similar at higher
values of $U$ as well.  This brings out clearly that $C_{60}$ is a
more frustrated system than its fragments. The benzene value is
clearly outside the range of the fragments of $C_{60}$.

In table (12) we present bond orders, spin--spin and charge--charge
correlations for $C_{60}$ with and without bond alternation, for the
h--h and h--p bonds as well as diagonal charge and spin correlations.
We notice that the h--h  bond orders are larger than the h--p bond
orders for undistorted $C_{60}$ (uniform bond length) and the
difference increases with alternation.  While it is possible to
optimise the bond length differences for the h--h and the h--p bonds
on the basis of Coulson's empirical formulae, we have not attempted
it. From the bond orders for the uniform transfer integral, Coulson's
formula\cite{coulson,mall1} predicts a h--h bond length of 1.429~$\AA$ and a h--p bond
length of 1.453~$\AA$.
However, we have chosen the transfer integral of the h--h bond to be
$20 \%$ stronger than that of
the uniform case and the h--p bond to be $10 \%$ smaller than in the
uniform case. This choice
conserves the total transfer integral for the buckyball and allows comparison with the
uniform  case. The larger alternation has been employed  to offset
the errors in monte carlo estimates. 

The difference in bond order between the h--h and h--p bonds in
uniform $C_{60}$ is larger  than that in corannulene.  This seems to
indicate that the bond frustrations in $C_{60}$ are more localized on
the pentagons than in the fragments of $C_{60}$ or indeed any open
frustrated systems.   The h--p and h--h charge--charge
correlations show the opposite trend when compared to corannulene. This
is because the charge--charge correlations are smaller for large bond
orders as the transfer of an electron requires the dissimilar
occupancy of the sites between  which hopping is occuring.  The
charge--charge correlation for the h--h bond is smaller than for the
h--p bond in $C_{60}$.  The
spin--spin correlation is more antiferromagnetic for the h--h bond
than for the h--p bond in $C_{60}$. The h--h bond in  $C_{60}$ has
weaker antiferromagnetic fluctuations than the h--h bond in corannulene,
while  the h--p bond in  $C_{60}$ has smaller antiferromagnetic fluctuations than
the h--p bond in corannulene. This trend holds for all values of $U$.

The site ionicities  of the carbon sites in $C_{60}$  are smaller
than the interior sites in corannulene, as can be seen from the smaller
diagonal charge correlation in the former.
This is also confirmed by the larger diagonal spin correlation in
$C_{60}$ compared with that for the pentagonal sites in corannulene. 

The spin and charge correlations for the h--h bond increase when
alternation in the transfer integral is introduced, while these
quantities decrease for the h--p bond. The site ionicities are
smaller than those of uniform $C_{60}$. This trend persists for
all values of $U$ studied by us. 

The average bond order of a molecule gives the extent of kinetic
stabilization due to delocalization.   It is interesting  to
compare the average bond orders for $U/t=0$  for benzene,
$C_{60}$ and graphite\cite{coul1}, which are 2/3,  0.518 and 0.525. In benzene 
and $C_{60}$ the bond orders are marginally  reduced for $U/t=1.0$  to
0.66 and 0.513 respectively.
A similar trend is expected in graphite, upon introducing
correlations.  Thus, it appears that in $C_{60}$ the kinetic stabilization
is only slightly smaller than in graphite, in spite of  bond frustration
in the former due to the five-membered rings. This result could
be an artefact of the Hubbard model; for even in  the mean-field limit,
extended range interactions renormalize the transfer integrals,
leading to contributions to  the kinetic energy and hence  to the
bond orders.
 It is also interesting to compare the pentagons in
$C_{60}$ with the cyclopentadienyl radical.  The pentagonal bond
order in this molecule is 0.58  while the h--p bond order in $C_{60}$ is
0.47 for $U=1.0$.  Thus the pentagons in $C_{60}$ are more frustrated
than a single pentagonal unit. 

The value of the  Gutzwiller parameter $g$ for which the energy is a
minimum indicates the overall extent of supression of the doubly
occupied sites in a given system. 
In table
(13), the value of the Gutzwiller parameter $g$ for which the energy
is a minimum is given for all the systems reported  in this paper.
There is no systematic variation in $g$ across the fragments and all
the systems seem to suppress double occupancies to a similar extent
for a given value of $U$.

\section{Summary}

We have studied the fragments of $C_{60}$ by various CI techniques as
well as the VMC. Comparison of these studies allows us to derive
reasonable monte carlo parameters for the simulation of larger
systems like $C_{60}$. The
bond order analysis of these molecules allows us to visualise
structures which are consistent with the spin and charge
correlations. We are able to observe the evolution of the properties
of the fragments towards the properties of $C_{60}$ through the
properties of the hexagon--hexagon like and hexagon--pentagon like bonds.
Corannulene is found to be  the fragment of $C_{60}$ that
approximates the full molecule most closely. We have studied the  uniform
and bond alternated models for  $C_{60}$. Even when the transfer integral is
uniform, the bond orders of $C_{60}$ show more marked alternation
than the fragments. 

Even though the VMC is an approximation over and above that inherent in
the Gutzwiller wavefunction it is interesting to see that
the VMC technique provides a reasonable description of  the ground
state electronic structure of molecular systems. The VMC energies are
in good agreement with model exact energies for intermediate values
of $U/t$, but progressively deteriorates for large $U/t$. We are currently
involved in extending the Gutzwiller type of  VMC approach to models
with nearest neighbour Coulomb interactions. 

\noindent
{\bf{Acknowledgements:}} We thank Y. Anusooya for assistance rendered in the MO--VB
computations on PPP models and Dr. Biswadeb Dutta for help
with the computational facility at JNCASR.
\pagebreak
\clearpage

\begin{table}
\begin{center}
Table 1. Ground state energy of benzene (in units of $t$) as a function of $U$ from
exact and approximate calculations. The $g_{min}$ values are given in
parantheses. The VMC energies do not change on doubling the number of MPE for
sampling \\
\begin{tabular}{ c| c c c c  } \\ \hline
 $U$ &   $E_{exact}$  & $E_{MO-VB}$ &  $E_{Gutz}$  &  $E_{VMC}$  \\ \hline
0.5 & -7.2744     & -7.2742  &  -7.2741 (0.91) & -7.274 (0.91)\\ 
1.0 & -6.6012     & -6.5963  &  -6.5964 (0.83) & -6.597 (0.83)\\
2.0 & -5.4105     & -5.3744  &  -5.3868 (0.69) & -5.388 (0.71)\\
4.0 & -3.6697     & -3.3516  &  -3.5371 (0.47) & -3.538 (0.46)\\ \hline
\end{tabular}
\end{center}
\end{table}

\begin{table}
\begin{center}
Table 2. Spin--Spin correlations ($4<s_{i}^{z}s_{j}^{z}>$) and
charge--charge correlations ($<n_{i}n_{j}>$) of benzene for $U = 1.0$
and 4.0, from
exact and approximate calculations. The VMC values do not change on
doubling the Monte Carlo sampling size. \\
\begin{tabular}{c|c|ccc|ccc} \\ \hline 
\multicolumn{1}{c|}{$U$}&\multicolumn{1}{c|}{$i,j$} & \multicolumn{3}{c|}
{$4<s_{i}^{z}s_{j}^{z}>$} & \multicolumn{3}{c|}{$<n_{i}n_{j}>$} \\ \hline 
 & &   $Exact$  & Gutz &VMC&   $Exact$  & Gutz &VMC  \\ \hline
1.0  &1,1 &  0.567     &  0.565 &0.557  &1.432 &1.435 &1.443 \\ 
1.0  &1,2 & -0.267     & -0.265 &-0.245 &0.816 &0.817 &0.812\\
1.0  &1,3 &  0.022     &  0.016 &0.011  &0.989 &0.989 &0.976\\
1.0  &1,4 & -0.077     & -0.068 &-0.062 &0.960 &0.953 &0.952\\ \hline
4.0  &1,1 & 0.778      &  0.751 & 0.750 &1.222 &1.249 &1.250\\
4.0  &1,2 &-0.435      & -0.404 &-0.395 &0.905 &0.911 & 0.911\\
4.0  &1,3 & 0.140      &  0.095 & 0.077 &0.999 &0.979 & 0.979\\
4.0  &1,4 &-0.188      & -0.131 &-0.113 &0.989 &0.971 & 0.971\\ \hline
\end{tabular}
\end{center}
\end{table}

\pagebreak
\clearpage

\begin{table}
\begin{center}
Table 3. Bond--orders of the nearest neighbour bond in benzene for  $U = 1.0$
and $4.0$ from exact and approximate calculations. The VMC values do not change on
doubling the Monte Carlo sampling size.\\
\begin{tabular}{c|ccc}  \\ \hline 
$U$  &   $Exact$  & Gutz    &VMC  \\ \hline
0.5  &  0.6646     & 0.6646 &0.664  \\
1.0  &  0.6582     & 0.6584 &0.660   \\ 
2.0  &  0.6316     & 0.6344 &0.636   \\
4.0  &  0.5278     & 0.5441 &0.545  \\ \hline
\end{tabular}
\end{center}
\end{table}

\begin{table}
\begin{center}
Table 4. Ground state energies of pyracylene, fluoranthene and corannulene (in units of $t$) from restricted CI and
VMC calculations. Numbers in parantheses are the energies per site.
MO--VB energies are not given for larger values of $U$ as the
technique is known to give poor ground state energies for large $U$.  
The VMC values do not change on
doubling the Monte Carlo sampling size.\\
\begin{tabular}{c|cc|cc|cc}\\ \hline 
\multicolumn{1}{c|}{$U$}&\multicolumn{2}{c|}{Pyracylene}&\multicolumn{2}{c|}
{Fluoranthene} &\multicolumn{2}{c|}{Corannulene} \\ \hline 
 & $E_{MO-VB}$ & $E_{VMC}$ &  $E_{MO-VB}$  &  $E_{VMC}$ &  $E_{MO-VB}$  &  $E_{VMC}$  \\ \hline 
0.5 &   -17.7002 & -17.713(-1.27) & -20.5231  & -20.558(-1.28) & -26.2689 &-26.306(-1.32)  \\
1.0 &   $-16.0631^{*}$ & -16.116(-1.15) & -18.5943  & -18.730(-1.17) & -23.8655&-24.007(-1.20) \\
2.0 &   -12.9817 & -13.235(-0.95) &-14.8631 &-15.422(-0.96) &-19.2221 & -19.847(-0.99)\\ 
4.0 &    ---    & -8.666 (-0.62) & ---    & -10.164 (-0.64)& ---&
-13.158(-0.66) \\ \hline
\end{tabular}
\end{center}
$^{*}$ exact energy obtained from full CI calculation in the singlet
subspace of dimension 2760615 is -16.13
\end{table}

\pagebreak
\clearpage

\begin{table}
\begin{center}
Table 5a. Bond-orders of the unique bonds and total kinetic energy
(KE) in the ground state of
pyracylene  from  VMC calculations. The VMC values do not change on
doubling the Monte Carlo sampling size. \\
\begin{tabular}{ c|ccccc } \\ \hline
$i,j$ &  $U=0.5$ & $U=1.0$ & $U=2.0$ & $U=4.0$ & $PPP(exact)$ \\ \hline
1,2   &  0.819 & 0.819 & 0.795 & 0.695  & 0.816 \\
2,3   &  0.399 & 0.402 & 0.382 & 0.330  & 0.239\\
3,4  &  0.517 & 0.509 & 0.494 & 0.442   & 0.455\\
5,6   &  0.652 & 0.647 & 0.630 & 0.562  & 0.697\\
6,9  &  0.609 & 0.605 & 0.581 & 0.518   & 0.604\\ 
4,11 &  0.531 & 0.531 & 0.519 & 0.463   & 0.526\\ \hline
KE  & -19.365 & -19.228 & -18.594 & -16.448 & -17.860 \\ \hline
\end{tabular}
\end{center}
\end{table}

\pagebreak
\clearpage

\begin{table}
\begin{center}
Table 5b. Normalized bond orders of the unique bonds in the ground state of
pyracylene  from  VMC calculations. \\
\begin{tabular}{ c|ccccc } \\ \hline
$i,j$ &  $U=0.5$ & $U=1.0$ & $U=2.0$ & $U=4.0$ & $PPP(exact)$ \\ \hline
1,2   &  1.000 &  1.000  &  1.000 & 1.000 & 1.000 \\ 
2,3   &  0.490 &  0.491  &  0.480 & 0.474 & 0.293 \\
3,4   &  0.631 &  0.621  &  0.621 & 0.636 & 0.558 \\
5,6   &  0.795 &  0.790  &  0.793 & 0.809 & 0.854 \\
6,9   &  0.790 &  0.739  &  0.731 & 0.745 & 0.740 \\
4,11  &  0.648 &  0.648  &  0.652 & 0.665 & 0.645 \\  \hline
\end{tabular}
\end{center}
\end{table}

\pagebreak
\clearpage

\begin{table}
\begin{center}
Table 6a. On--site and nearest neighbour VMC spin--spin correlations ($4<s_{i}^{z}s_{j}^{z}>$) 
 of pyracylene for different $U/t$ values. The VMC values do not change on
doubling the Monte Carlo sampling size.  Numbers in parantheses are the
exact AO-VB results, given for comparison.\\
\begin{tabular}{cccc|cccc} \\ \hline 
\multicolumn{4}{c|}{Nearest neighbour} &\multicolumn{4}{c|}{On--site} \\ \hline 
$i,j$ & $U=1.0$ & $U=2.0$ & $U=4.0$ & $i,j$  &  $U=1.0$& $U=2.0$& $U=4.0$  \\ \hline 
1,2 &   -0.402 (-0.4308) & -0.475 & -0.595 &1,1& 0.565(0.5849)&0.625&0.744 \\
2,3 &   -0.099 (-0.0930) & -0.107 & -0.138 &3,3& 0.551(0.5563)&0.616&0.711 \\
3,4 &  -0.155 (-0.1532) & -0.182 & -0.241 &4,4& 0.549 (0.5477)&0.604&0.706 \\
5,6 &   -0.250 (-0.2598) & -0.254 & -0.382&6,6& 0.554 (0.5651)&0.617  & 0.719 \\
4,11 & -0.164 (-0.1703) & -0.205 & -0.273 \\
6,9 &  -0.225 (-0.2219) & -0.260 & -0.326 \\ \hline
\end{tabular}
\end{center}
\end{table}

\pagebreak
\clearpage

\begin{table}
\begin{center}
Table 6b. On--site and nearest neighbour VMC charge--charge
correlations  ($<n_{i}n_{j}>$) 
 of pyracylene for different $U/t$ values.  The VMC values do not change on
doubling the Monte Carlo sampling size.\\
\begin{tabular}{cccc|cccc} \\ \hline 
\multicolumn{4}{c|}{Nearest neighbour} &\multicolumn{4}{c|}{On--site} \\ \hline 
$i,j$ & $U=1.0$ & $U=2.0$ & $U=4.0$ & $i,j$  &  $U=1.0$& $U=2.0$& $U=4.0$  \\ \hline 
1,2 &   0.794   &  0.841 & 0.929 &1,1 &   1.514   &  1.444 & 1.321 \\
2,3 &   0.964   &  0.972 & 0.992 &3,3 &   1.431   &  1.374 & 1.281 \\
3,4 &   0.964   &  0.984 & 1.012 &4,4 &   1.621   &  1.570 & 1.400 \\
5,6 &   0.737   &  0.776 & 0.846 &6,6 &   1.228   &  1.240 & 1.171 \\ 
6,9 &   0.702   &  0.738 & 0.810 & & & &\\
4,11&   1.059   &  1.064 & 1.393 & & & &\\ \hline
\end{tabular}
\end{center}
\end{table}

\pagebreak
\clearpage

\begin{table}
\begin{center}
Table 7. Bond-orders of the unique bonds and total kinetic energy
(KE) in the ground state of
fluoranthene  from  VMC calculations.  The VMC values do not change on
doubling the Monte Carlo sampling size. \\
\begin{tabular}{ c|ccccc } \\ \hline
$i,j$ &   $U=1.0$ & $U=2.0$ & $U=4.0$  \\ \hline
1,2  &    0.643 & 0.612  &  0.518  \\
2,3  &    0.666 &  0.637  &  0.550  \\
3,4  &    0.611 &  0.591  &  0.519  \\
4,5  &    0.388 &  0.377  &  0.320  \\
5,6  &    0.657 &  0.630  &  0.548  \\
6,7  &    0.611 &  0.593  &  0.514  \\
7,8  &    0.707 &  0.676  &  0.579  \\
8,9  &    0.546 &  0.522  &  0.451  \\
4,14 &    0.566 &  0.539  &  0.472  \\
5,16 &    0.512 &  0.494  &  0.436  \\
9,16 &    0.526 &  0.512  &  0.446  \\ \hline
KE   &  -22.256 & -21.397 & -18.539  \\ \hline
\end{tabular}
\end{center}
\end{table}
\pagebreak
\clearpage

\begin{table}
\begin{center}
Table 8a. On--site and nearest neighbour VMC spin--spin correlations ($4<s_{i}^{z}s_{j}^{z}>$) 
 of fluoranthene for different $U/t$ values. The VMC values do not change on
doubling the Monte Carlo sampling size. \\
\begin{tabular}{cccc|cccc} \\ \hline 
\multicolumn{4}{c|}{Nearest neighbour} &\multicolumn{4}{c|}{On--site} \\ \hline 
$i,j$ & $U=1.0$ & $U=2.0$ & $U=4.0$ & $i,j$  &  $U=1.0$& $U=2.0$& $U=4.0$  \\ \hline 
1,2   &  -0.259  &   -0.300  &  -0.395  &1,1& 0.5617& 0.6358& 0.7548  \\
2,3   &  -0.284  &   -0.327  &  -0.409 &3,3 & 0.5702& 0.6372&   0.7399  \\
3,4   &  -0.228  &   -0.283  &  -0.362 &4,4   &   0.552  &    0.619  &   0.729  \\
4,5   &  -0.087  &   -0.109  &  -0.137 &5,5   &   0.561  &    0.622  &   0.729  \\
5,6   &  -0.260  &   -0.305  &  -0.403 &6,6   &   0.552  &    0.621  &   0.738  \\
6,7   &  -0.222  &   -0.271  &  -0.360 &7,7   &   0.565  &    0.630  &   0.745  \\
7,8   &  -0.303  &   -0.361  &  -0.452 &8,8   &   0.557  &    0.626  &   0.744  \\
8,9   &  -0.176  &   -0.206  &  -0.280 &9,9   &   0.551  &    0.620  &   0.731  \\
4,14  &  -0.187  &   -0.220  &  -0.298 &16,16 &   0.561  &    0.624  &   0.722  \\ 
5,16  &  -0.167  &   -0.211  &  -0.271  & & & &\\
9,16  &  -0.166  &   -0.176  &  -0.233 & & & & \\ \hline
\end{tabular}
\end{center}
\end{table}

\pagebreak
\clearpage

\begin{table}
\begin{center}
Table 8b. On--site and nearest neighbour VMC charge--charge
correlations  ($<n_{i}n_{j}>$) 
 of fluoranthene for different $U/t$ values. \\
\begin{tabular}{cccc|cccc} \\ \hline 
\multicolumn{4}{c|}{Nearest neighbour} &\multicolumn{4}{c|}{On--site} \\ \hline 
$i,j$ & $U=1.0$ & $U=2.0$ & $U=4.0$ & $i,j$  &  $U=1.0$& $U=2.0$& $U=4.0$  \\ \hline 
1,2   &   0.835  &    0.874  &   0.926 &1,1   &   1.452  &    1.373  &   1.249  \\
2,3   &   0.807  &    0.859  &   0.911 &3,3   &   1.397  &    1.364  &   1.260  \\
3,4   &   0.864  &    0.898  &   0.945 &4,4   &   1.523  &    1.433  &   1.318  \\
4,5   &   1.013  &    1.003  &   1.016 &5,5   &   1.515  &    1.443  &   1.321  \\
5,6   &   0.800  &    0.841  &   0.909 &6,6   &   1.344  &    1.291  &   1.209  \\
6,7   &   0.790  &    0.825  &   0.883 &7,7   &   1.435  &    1.373  &   1.234  \\
7,8   &   0.839  &    0.849  &   0.861 &8,8   &   1.379  &    1.294  &   1.193  \\
8,9   &   0.839  &    0.860  &   0.914 &9,9   &   1.452  &    1.384  &   1.278  \\
4,14  &   0.934  &    0.939  &   0.981 &16,16 &   1.508  &    1.452  &   1.343  \\ 
5,16  &   0.965  &    0.986  &   0.998 & & & & \\
9,16  &   0.920  &    0.949  &   0.981 & & & & \\ \hline
\end{tabular}
\end{center}
\end{table}

\pagebreak
\clearpage

\begin{table}
\begin{center}
Table 9. Bond-orders of the unique bonds and total kinetic energy
(KE) in the ground state of
corannulene  from  VMC calculations. \\
\begin{tabular}{ c|ccc } \\ \hline
$i,j$ &   $U=1.0$ & $U=2.0$ & $U=4.0$   \\ \hline
1,2  &    0.745 &  0.721  &  0.641  \\
2,3  &    0.529 &  0.509  &  0.454  \\
16,17  &  0.504 &  0.489  &  0.433  \\
3,17  &   0.551 &  0.535  &  0.491  \\
KE   &   -28.562 & -27.673 & -24.751  \\ \hline
\end{tabular}
\end{center}
\end{table}

\begin{table}
\begin{center}
Table 10a. On--site and nearest neighbour VMC spin--spin correlations ($4<s_{i}^{z}s_{j}^{z}>$) 
 of corannulene for different $U/t$ values. \\
\begin{tabular}{cccc|cccc} \\ \hline 
\multicolumn{4}{c|}{Nearest neighbour} &\multicolumn{4}{c|}{On--site} \\ \hline 
$i,j$ & $U=1.0$ & $U=2.0$ & $U=4.0$ & $i,j$  &  $U=1.0$& $U=2.0$& $U=4.0$  \\ \hline 
1,2  &  -0.323 & -0.380  & -0.411 & 1,1 & 0.552 & 0.612 & 0.720 \\
2,3  &  -0.155 & -0.194  & -0.240 & 3,3 & 0.539 & 0.604 & 0.702 \\
16,17 & -0.148 & -0.183  & -0.212 & 16,16 & 0.537 & 0.601 & 0.699 \\
3,17 &  -0.173 & -0.211  & -0.278 \\ \hline
\end{tabular}
\end{center}
\end{table}
\pagebreak

\pagebreak
\clearpage

\begin{table}
\begin{center}
Table 10b. On--site and nearest neighbour VMC charge--charge
correlations  ($<n_{i}n_{j}>$) 
 of corannulene for different $U/t$ values. \\
\begin{tabular}{cccc|cccc} \\ \hline 
\multicolumn{4}{c|}{Nearest neighbour} &\multicolumn{4}{c|}{On--site} \\ \hline 
$i,j$ & $U=1.0$ & $U=2.0$ & $U=4.0$ & $i,j$  &  $U=1.0$& $U=2.0$& $U=4.0$  \\ \hline 
1,2  &  0.725 & 0.777  & 0.847 & 1,1 & 1.414 & 1.359 & 1.251 \\
2,3  &  0.840 & 0.870  & 0.915 & 3,3 & 1.422 & 1.357 & 1.284 \\
16,17  &  0.997 & 1.001  & 1.017 & 16,16 & 1.567 & 1.494 & 1.378 \\
3,17  &  0.895 & 0.917  & 0.957 \\ \hline
\end{tabular}
\end{center}
\end{table}
\pagebreak

\begin{table}
\begin{center}
Table 11. Ground state energy of uniform and bond alternant $C_{60}$ for various values of $U$
from VMC.\\
\begin{tabular}{c|ccc|ccc} \hline 
\multicolumn{1}{c|}{$U$} & \multicolumn{3}{c|}{uniform}&
\multicolumn{3}{c|}{bond alternant} \\ \hline 
    & $E$ & $E/site$ & $E/bond$  & $E$  & $E/site$ & $E/bond$  \\ \hline 
1.0 & -78.891  & -1.315 & -0.877  & -81.448 & -1.358 & -0.905\\
2.0 & -66.074  & -1.101 & -0.734  & -68.733 & -1.146 & -0.764\\
3.0 & -55.669  & -0.928 & -0.619  & -57.473 & -0.958 & -0.639\\
4.0 & -44.678  & -0.745 & -0.496  & -47.768 & -0.796 & -0.531\\
5.0 & -36.088  & -0.602 & -0.401  & -39.566 & -0.695 & -0.440\\
6.0 & -28.787  & -0.480 & -0.320  & -32.435 & -0.541 & -0.360\\  \hline 
\end{tabular}
\end{center}
\end{table}

\pagebreak
\clearpage

\begin{table}
\begin{center}
Table 12. On--site and nearest neighbour VMC charge--charge
correlations  ($<n_{i}n_{j}>$) and spin--spin correlations
($4<s_{i}^{z}s_{j}^{z}>$) and bond orders 
 of uniform and bond alternant $C_{60}$ for different $U/t$ values. \\
\begin{tabular}{c|c|ccc|ccc} \\ \hline 
\multicolumn{1}{c|}{$U$} &\multicolumn{1}{c|}{}&
\multicolumn{3}{c|}{uniform}& \multicolumn{3}{c|}{bond alternant} \\ \hline 
 &  & h--h & h--p & diag  &  h--h& h--p& diag  \\ \hline 
1.0  & bo & 0.597 & 0.471  &---  & 0.849 & 0.367 & --- \\
  &  s-s  &-0.212  & -0.132  & 0.550 & -0.296 & -0.099 & 0.5500 \\
  &  c-c  & 0.847  & 0.904  & 1.450 & 0.784 & 0.927 & 1.450 \\ \hline
2.0 &bo &  0.582 & 0.461  &---  & 0.827 & 0.359 & --- \\
  & s-s& -0.241 & -0.152  & 0.589 & -0.336 & -0.115 & 0.601 \\
  & c-c & 0.872 & 0.919  & 1.402 & 0.819 & 0.938 & 1.399 \\ \hline
3.0& bo  & 0.535 & 0.414  &---  & 0.756 & 0.321 &---  \\
  & s-s & -0.314 & -0.191  & 0.690 & -0.431 & -0.143 & 0.699 \\
 & c-c &0.912 & 0.947 & 1.310 & 0.879 & 0.960 & 1.301 \\ \hline
4.0 & bo& 0.448 & 0.336 &---  & 0.634 & 0.254 &---  \\
 & s-s &-0.404 & -0.233 & 0.787 & -0.545 & -0.169 & 0.799 \\
 & c-c& 0.949 & 0.970 & 1.258 &0.928 &0.977 & 1.201 \\ \hline 
\end{tabular}
\end{center}
\end{table}
\pagebreak
\clearpage

\begin{table}
\begin{center}
Table 13. Minimum value of the Gutzwiller parameter $g$ for fragments
of $C_{60}$, uniform and bond alternant $C_{60}$ (A and B). \\
\begin{tabular}{ c|ccccc } \\ \hline
$U$ & pyracylene & fluoranthene & corannulene &$C_{60}$(A) &$C_{60}$(B)    \\ \hline
0.5 & 0.91  & 0.92 & 0.93 &--- &--- \\
1.0 & 0.85 & 0.83 & 0.88 & 0.85 & 0.87  \\
2.0 & 0.69  & 0.69 & 0.71 & 0.73& 0.74 \\
4.0 & 0.50  & 0.47 & 0.51& 0.53 &0.53\\ \hline
\end{tabular}
\end{center}
\end{table}

\pagebreak

\pagebreak
\clearpage
\begin{center}
{\bf \underline{Figure Captions}} \\
\end{center}

\begin{description}
{\item {\bf Figure 1.}} Structures of Pyracylene, Fluoranthene and Corannulene. 
{\item {\bf Figure 2.}}  Bond orders(A), diagonal charge--charge
(spin--spin) correlations(B), narest neighbour charge--charge
correlations(C) and nearest neighbour spin--spin correlations(D) for
unique sites and bonds of
pyracylene for $U=1.0$.
{\item {\bf Figure 3a.}}  Bond orders and  diagonal charge--charge
(spin--spin) correlations for unique bonds and sites in fluoranthene
for $U=1.0$. The axes of symmetry are shown. 
{\item {\bf Figure 3b.}}  Unique nearest neighbour charge--charge
correlations and  spin--spin correlations of
fluoranthene for $U=1.0$.
{\item {\bf Figure 4.}} Bond orders and  diagonal charge--charge
(spin--spin) correlations of in corannulene for  $U=1.0$. The
difference between  symmetry related quantities indicates the extent
of scatter in VMC estimates. 
{\item {\bf Figure 5.}} Nearest neighbour charge--charge
correlations and  spin--spin correlations  in corannulene for $U=1.0$. 
The difference between  symmetry related quantities indicates the extent
of scatter in VMC estimates. 
\end{description}

\end{document}